\documentclass[aps,prb,twocolumn,showpacs,preprintnumbers,amsmath,amssymb,amsfonts,floatfix]{revtex4}

\usepackage{epstopdf}
\usepackage{subfigure}
\usepackage[latin1]{inputenc}
\usepackage{wasysym}
\usepackage{color}
\usepackage{dcolumn}
\usepackage{dcolumn}
\usepackage{longtable}
\usepackage{graphicx}
\usepackage{amssymb}
\usepackage{dcolumn}
\usepackage{bm}
\usepackage{bbm}
\usepackage{mathrsfs}

\newcommand{\beq}{\begin{eqnarray}}
\newcommand{\eeq}{\end{eqnarray}}
\newcommand{\bem}{\begin{pmatrix}}
\newcommand{\eem}{\end{pmatrix}}
\newcommand{\nn}{\nonumber}
\newcommand{\hb}{\hbar}

\newcommand{\f}{\frac}

\newcommand{\dr}[1]{|#1\rangle}
\newcommand{\dl}[1]{\langle#1}

\newcommand{\tr}[1]{\textrm{#1}}

\def\nn{\nonumber}

\begin{document}
\title{Theoretical analysis of the density of states of graphene at high magnetic field using Haldane pseudopotentials}
\author{Lih-King Lim$^{1}$}
\author{M. O. Goerbig$^{1}$}
\author{Cristina Bena$^{1,2}$}
\affiliation{$^1$Laboratoire de Physique des Solides, CNRS UMR 8502, Univ. Paris-Sud, F-91405 Orsay cedex, France\\
$^2$Institut de Physique Th\'{e}orique, CEA/Saclay, CNRS, URA 2306, Orme des Merisiers, F-91191 Gir-sur-Yvette, France}
\date{\today}

\begin{abstract}
We study the density of states in graphene at high magnetic field, when the physics is dominated by strong correlations between electrons. In particular we use the method of Haldane pseudopotentials to focus on almost empty or almost filled Landau levels. We find  that, besides the usual Landau level peaks, additional peaks (``sashes'') appear in the spectrum. The energies of these peaks are determined by the strength of Haldane's pseudopotentials, but as opposed to the usual two-dimensional gas, when there is a one-to-one correspondence between a Haldane pseudopotential and a peak in the spectrum, the energy of each peak is determined in general by a combination of more than one pseudopotential values.  An eventual measure of these peak in the density of states spectrum of graphene would allow one to determine the value of the pseudopotentials in graphene, and thus test the strength of the interactions in this system.
\end{abstract}

\maketitle

\section{Introduction}

As high quality samples of graphene are now readily available,\cite{Morozov:08,Du:08,Bolotin:08} an exciting prospect is to observe signatures of strong electron correlations in this system. Graphene's distinctive feature with respect to a more conventional two-dimensional (2D) electron gas such as the one realized in GaAs-based structures is that its charge carriers are relativistic in character, having a linear energy dispersion.\cite{Geim:07, Neto:09} From a theoretical perspective it is interesting to understand the interplay between the Dirac character of the carriers and the effects of interactions between the electrons.\cite{Peres:11,Goerbig:10,Kotov:10} While the linearity of the spectrum has already been demonstrated by different experiments,\cite{Neto:09} the effects of interactions have been believed to be quite small in standard graphene samples. Recently, however, the fractional quantum Hall effect has been observed in suspended graphene,\cite{Du:09,Bolotin:09} demonstrating that interactions can play an important role in graphene in a strong magnetic field with sufficiently high mobility.

The analysis of fractional quantum Hall (FQH) states arising at high perpendicular magnetic field in a two-dimensional system has revealed most prominently the effects of electronic interactions.\cite{Tsui:82,Laughlin:83,Pinczuk:96} In GaAs-based 2D electron gases in the FQH regime important information has been obtained from transport and shot noise experiments.\cite{Picciotto:97,Saminadaya:97} However, it is known that various 2D correlated states do not show anomalous transport signatures.\cite{Jain:89} Moreover, due to the manner of construction of the quantum well structures, having a direct access to their bulk properties is in general not straightforward. For example, only a few measurements of the local density of states  have been performed in 2D gases,\cite{Chan:97,Dial:07,hashim:08}, in general via indirect methods.

Compared to the other 2D gases, graphene has the advantage that being a surface electronic system, in addition to standard transport measurements, it is directly accessible by local density of states (DOS) measurements such as scanning tunneling microscopy (STM).\cite{Li} This opens the perspective to use such measurements to obtain information about the electronic interactions. Here we study the DOS in graphene exposed to a strong magnetic field focusing on the effects of the electronic interactions, taken into account via the method of Haldane pseudopotentials.\cite{Haldane:83}

Our analysis has been motivated by a recent experimental analysis of the high-resolution time-domain capacitance spectroscopy for a 2D quantum well in the quantum Hall regime.\cite{Dial:07} The observation of unexpected peaks in the high-energy spectra for LL filling factors near $\nu=0$ and $\nu=1$, to which the authors referred as ``sashes", has called for new perspectives in the problem of 2D correlation physics in strong magnetic fields. The filling factor $\nu=n_{el}/n_B$ is the ratio between the electronic density $n_{el}$ and that $n_B=eB/h$ of flux quanta threading the 2D surface of the system. Soon after the experimental observation, a few different approaches have been proposed for the description of these peaks in relation to the electronic interactions.\cite{MacDonald:10,Barak:10} In particular, in Ref.~\onlinecite{MacDonald:10}  these peaks have been attributed to the strong correlations between electrons, which have been modeled using the Haldane's pseudopotentials.

In the present paper, we generalize this calculation for 2D electrons in graphene. In this case, the total filling factor is defined as $\nu=\nu_n+\bar{\nu}$, where $\nu_n=\pm 2(2n+1)=\pm2,\pm6,\ldots$ is the filling at which relativistic integer quantum Hall effect occurs\cite{Peres:11,Goerbig:10} and $\bar{\nu}$ is the partial filling factor of the $n$-th LL. In this approach, at very low partial filling factors $\bar{\nu}\rightarrow 0$, besides a number of completely filled and inert LLs, only one extra particle is considered present in the ground state. Whereas this theoretical limit allows for an important simplification in the calculation of the tunneling density of states, it is experimentally relevant as long as the average distance between particles $d\sim l_B/\sqrt{\bar{\nu}}$ of electrons in the $n$-th LL is larger than the cyclotron radius $R_C=l_B\sqrt{2n+\delta_{n,0}}$ in graphene, in terms of the magnetic length $l_B=\sqrt{\hbar c/eB}\simeq 26\,\tr{nm}/\sqrt{B[T]}$. Our theoretical analysis is therefore applicable in the limit
$\bar{\nu}\ll 1/(2n+\delta_{n,0})$.

Assuming thus that the state of the system is a one-particle state, when a second particle tunnels in from the STM tip, the measured STM signal is proportional to the overlap between the resulting state, which is not an eigenstate of the high-magnetic field two-particle Hamiltonian, and the two-particle eigenfunctions. These eigenfunctions are characterized by a quantum number associated with the two-particle relative angular momentum. As a result, discrete peaks arise in the density of states \cite{lectures} corresponding to the energy  differences between the two-body interacting eigenfunctions of the system and the one-particle state. These energy differences are given by Haldane's pseudopotentials, and their measurement via the peaks in the spectrum yields information about the interaction parameters in the system.

Similarly to non-relativistic 2D electron systems, such as in GaAs heterostructures, we find that the high-field DOS, close to the filling $\nu_n=\pm 2(2n+1)$, allows for a determination of Haldane's pseudopotentials and thus of the effective Coulomb interaction in graphene. However, contrary to
non-relativistic 2D electrons in a strong magnetic field, the center-of-mass (CM) is not separable from the relative coordinate as
a consequence of the Lorentz invariance. The extraction of Haldane's pseudopotentials in higher LLs ($n\neq 0$) is therefore more
involved than in $n=0$, where the two-body problem is equivalent to that in non-relativistic 2D electron systems.
Our results may be tested in high-field STM that has already been applied successfully in the past to graphene\cite{highBSTM},
and which has also been proposed as a tool for the study of high-field electron-solid phases\cite{popl}, as well as of the role of impurity scattering.\cite{bena:08,bena:10,cheianov:06}

The paper is organized as follows: in Sec. \ref{sec:2}, we review the recent theoretical interpretation of the ``sash" features observed in high-quality 2DEG samples via Haldane's pseudopotentials. In Sec. \ref{sec:3}, we generalize this theory to graphene by solving the two-body problem for Dirac quasiparticles in a strong magnetic field. With the help of the exact two-body eigenstates of the interacting system, in Secs. \ref{sec:4} and \ref{sec:5} we calculate the DOS, and we describe how it can be used to extract information about the pseudopotentials for the LLs $n=0$ and $n\neq 0$, respectively.  The last section (Sec. \ref{sec:6}) of the paper presents a discussion of the results and the conclusions.

\section{Theoretical interpretation of the DOS under high magnetic field using  Haldane pseudopotentials}
\label{sec:2}

Here we review the main aspects of  the theory based on Haldane's pseudopotentials proposed in Ref.~\onlinecite{MacDonald:10} to explain the unexpected sashes that have been recently observed in the DOS of a 2D gas at high magnetic field.\cite{Dial:07}  In standard STM experiments, the measured differential conductance is taken to be proportional to the tunneling DOS of the system being probed. At zero temperature, this quantity is given by
\beq
A(\omega)= \sum_{\alpha} |\langle \Psi_{\alpha}(N+1)\bigl|c_{\beta}^\dag|\Psi_0(N)\rangle\bigr|^2 \delta\bigl(\omega-E_{\alpha,0}\bigr)\nn\\
\tr{\ }+\sum_{\alpha}\bigl|\langle \Psi_{\alpha}(N-1)|c_{\beta}|\Psi_0(N)\rangle\bigr|^2 \delta\bigl(\omega+E_{\alpha,0}\bigr),
\eeq
where $|\Psi_0(N)\rangle$ is the ground state of the $N$-particle system and $c_{\beta}^{(\dagger)}$ is a fermionic operator
that removes (adds) a particle from (to) the one-particle state labeled by a set of quantum numbers $\beta$. In the high-field case for graphene, which we consider here, the set of quantum numbers is given by $\beta=\{n,m,\sigma,K/K'\}$, where $n$ is the LL index, $m$ labels the degenerate one-particle states inside the level, $\sigma$ is the spin index and $K/K'$ is the valley index.\cite{lectures} Thus, the first term describes an electron added to the system, whereas the second one corresponds to removing an electron. Here, $|\Psi_{\alpha}(N\pm1)\,\rangle$ are the exact eigenstates of the interacting $N\pm1$-particle system labeled by another set of quantum numbers $\alpha$. Their energy difference is given by $E_{\alpha,0}=[E_{\alpha}(N\pm1)-E_0(N)]$. Because the quantum number $m$ is associated with the center of the electronic cyclotron motion, which is a constant of motion, the DOS is independent of $m$ in a translationally invariant system.\cite{lectures} Since the DOS yields information about the spectral properties of the many-body system, STM experiments can thus provide important information on the correlation physics in the quantum Hall regime.

To evaluate the above formula, in Ref.~\onlinecite{MacDonald:10}, one considers the extremely dilute limit $(\bar{\nu}\simeq0)$, where the ground state simply consists of a single particle in the lowest LL $|\Psi_0(N=1)\,\rangle$. Due to particle-hole symmetry, these results also hold for the electron-removal part of the DOS when the system is close to an almost filled LL state. The tunneling experiment in this regime is then described by adding an extra particle instantaneously to the ground state (without perturbing it). The resulting state is not an eigenstate of the Hamiltonian, the exact two-body eigenstates $|\Psi_{\alpha}(N=2)\,\rangle$ of a 2D gas in a strong magnetic field being labeled by their relative angular momentum. Moreover, the eigenvalues corresponded to the two-particle interacting states are shifted by the Coulomb interactions, the corresponded shifts being described by the Haldane's pseudopotentials.\cite{Haldane:83}. The resulting spectral peaks in the DOS spectrum occur at energies matching the difference between the energy of the interacting two-particle states and that of of the one-particle states -- this difference being precisely given by the Haldane's pseudopotentials. Thus the DOS spectrum provides detailed information about the values of the pseudopotentials in a system, which is crucial for understanding the formation of various correlated ground states.

\section{Two-body eigenstates of Dirac particles under magnetic field}
\label{sec:3}

As discussed in the previous section, knowledge of the exact two-body interacting eigenstates and their corresponding eigenvalues, as well as of the non-interacting state resulting by the addition of an extra particle to the one-particle state,  is required in order to calculate the DOS in the low-filling-factor limit. Given that the  two-particle eigenstates of the non-interacting Hamiltonian are also eigenstates of the Coulomb interaction potential, the two-particle interacting eigenstates have the same form as the non-interacting ones, but correspond to different eigenvalues. They are slightly more complicated for graphene for which the relevant charge excitations are Dirac-like particles, than for a 2D gas. Besides, one needs to take into account the extra spin and valley degrees of freedom. To begin, in Sect. IIIA, we focus on the orbital part of the interacting wavefunction for two Dirac particles in a magnetic field. In Sect. IIIB we write down the wavefunction obtained when adding an extra Dirac particle to a one-particle state. The effects of the extra degrees of freedom (spin, valley) will be touched upon in Sect. IIIC

\subsection{Two-particle interacting eigenstates and the corresponding eigenvalues}
As mentioned above the two-particle interacting eigenstates are the same as the non-interacting ones. In order to calculate them, one notes that at low-energy (up to fractions of an $eV$), the electronic properties of graphene can be described using a continuum model, with electrons localized around the two Dirac cones, conventionally called the $K$ and $K'$ valleys.\cite{Neto:09} The Schr\"{o}dinger equation for a Dirac particle in the $K$ valley in a magnetic field can be written as
\beq
H \Psi = v\vec{\sigma}\cdot\vec{\Pi}\,\Psi=E\Psi,
\eeq
where the Fermi velocity $v\simeq 10^6$ m/s plays the role of the speed of light. Furthermore, $\vec{\sigma}=(\sigma^x,\sigma^y)$ in terms of Pauli matrices, and $\vec{\Pi}=(\Pi_x,\Pi_y)$ is the canonical momentum operator, after Peierls substitution, which obeys $[\Pi_x,\Pi_y]=-i\hb^2/l_B^2$. 
By defining the ladder operators $a=l_B(\Pi_x-i\Pi_y)/\sqrt{2}\hb$ and $a^\dag=l_B(\Pi_x+i\Pi_y)/\sqrt{2}\hb$ that obey the harmonic oscillator algebra, $[a,a^\dag]=1$, the Schr\"{o}dinger equation becomes
\beq
\sqrt{2}\hb \frac{v}{l_B}\bem 0&a\\a^\dag&0\eem\Psi=E\Psi.
\eeq
Diagonalizing the Hamiltonian yields the energy spectrum of relativistic LLs
\beq
E=\lambda \hb \frac{v}{l_B}\sqrt{2 n},
\eeq
where $\lambda=\pm$, and $n=0,1,\ldots$ denotes the LL. The associated eigenstates are given by
\beq
&&\Psi_{n=0,m}=\bem 0 \\ |n=0,m\rangle \eem,\  \text{for}~n=0\nn\\
&&\Psi_{\lambda n,m}=\f{1}{\sqrt{2}}\bem |n-1,m\rangle \\\lambda |n,m\rangle \eem,\  \text{for}~n\neq 0,
\eeq
in terms of the (non-relativistic) LL states  $|n,m\rangle=\f{(a^\dag)^n (b^\dag)^m}{\sqrt{n!m!}}|0\rangle$. Here, we have implicitly introduced the cyclotron-orbit-center operator $b=[x+iy +\f{i}{m\omega}(\Pi_x+i\Pi_y)]$ that has the associated quantum number $m$ with $b^\dag b \ \mathbb{I}\  \Psi_{n,m} = m \Psi_{n,m} $, where $\mathbb{I}$ is the $2\times 2$ identity matrix. Since $[a,b]=[a^\dag,b]=0$, and $[H, b^\dag b \ \mathbb{I}]=0 $, the additional quantum number $m$ labels the macroscopic degeneracy $N_{B}=n_B\mathcal{A}$ for each Landau level $n$, in terms of the total surface $\mathcal{A}$.

The solutions of the same problem for the second valley $K'$ can be obtained by the transformation $\Psi\rightarrow \sigma_x \Psi$, since the the Hamiltonian around the $K'$ valley is related to Eq.~(2) via $H\rightarrow \sigma_x H \sigma_x^\dag$. This simple covariance property permits to focus the discussion in the rest of this only on one $K$ valley (assuming that the interaction is invariant under the above transformation, for more details see Sect. IIIC).

For the two-body problem of Dirac particles in the absence of magnetic field, it was shown \cite{Sabio:10} that due to the coupling of sublattice and orbital degrees of freedom, the CM degree of freedom cannot in general be separated from the relative coordinates degree of freedom. In the presence of a magnetic field, we have
\beq
v\bigl[ \sigma\cdot\vec{\Pi}_1\otimes \mathbb{I}+\mathbb{I}\otimes\sigma\cdot\vec{\Pi}_2\bigr]\Psi&=&E_0\,\Psi,
\eeq
or explicitly
\beq
\biggl(\f{\sqrt{2}\hb v}{l_B}\biggr) \bem 0&a_2&a_1&0\\ a_2^\dag&0&0&a_1\\ a_1^\dag&0&0&a_2\\ 0&a_1^\dag&a_2^\dag&0\eem\Psi&=&E_0\,\Psi,
\eeq
where the index $1,2$ labels the two particles. By defining the CM and relative coordinates, $z_R=(z_1+z_2)/2$ and $z_r=z_1-z_2$, respectively, the ladder operators become $a_R=( a_1+a_2)/\sqrt{2}, \tr{\ } a_r=( a_1-a_2)/\sqrt{2}$. The Schr\"{o}dinger equation then takes the form
\beq
 \biggl(\f{\hb v}{l_B}\biggr)
\bem 0&a_R-a_r&a_R+a_r&0\\ a_R^\dag-a_r^\dag&0&0&a_R+a_r\\ a_R^\dag+a_r^\dag &0&0&a_R-a_r\\ 0&a_R^\dag+a_r^\dag&a_R^\dag-a_r^\dag&0\eem\Psi=E_0\,\Psi,\nn\\
\eeq
which shows that this is also the case in the presence of a magnetic field.

We denote
\beq
\dr{N,M}_R\dr{n,m}_r\equiv\f{(a_R^\dag)^N (a_r^\dag)^{n} (b_R^\dag)^M (b_r^\dag)^{m}}{\sqrt{N!n!M!m!}}|0,0\rangle_R|0,0\rangle_r,\nn\\
\eeq
where we have defined $b_R=( b_1+b_2)/\sqrt{2}, \tr{\ } b_r=( b_1-b_2)/\sqrt{2}$ as the LL ladder operators, with the subscripts $R$ and $r$ in $\dr{N,M}_R\dr{n,m}_r$ indicating the CM and relative coordinates subspaces. Using this notation one may write down the eigenstates in a  4-spinor form $\dr{\vec{\Psi}_{M,m}(N=2)\,}$, see Table~\ref{table1} for the exact form of the lowest energy eigenstates.  We remark that the macroscopic degeneracies in both the CM and relative angular momenta, $M$ and $m$, are similar to the single-particle LL case.

While the eigenstates for the non-interacting and interacting problems are the same, the values of the eigenvalues are different. In order to compute these values, we now take into account the interactions
\beq
\bigl[ v\sigma\cdot\vec{\Pi}_1\otimes \mathbb{I}+\mathbb{I}\otimes v\sigma\cdot\vec{\Pi}_2+\hat{V}_{1,2}\mathbb{I}\otimes\mathbb{I}
\bigr]\Psi &=&E\,\Psi,
\eeq
where we consider the interaction potential to be isotropic $\dl r_1,r_2| \hat{V}_{1,2}\dr{r_1,r_2}= V(|r_1-r_2|)$. The eigenvalues $E$ of  the fully interacting Hamiltonian in Eq.~(10) can be obtained by sandwiching it between the eigenstates $\dr{\vec{\Psi}_{M,m}(N=2)\,}$ described above, with the help of Haldane's pseudopotentials,\cite{lectures}
\beq
V_{m}^{n}\equiv _{\ r}\!\!\langle n,m| \hat{V}_{1,2} |n,m \rangle_r.
\eeq
These eigenvalues are summarized in the third column of Table~\ref{table1}.
We can see that the interaction partially lifts the degeneracy in the relative angular momentum quantum number $m$ within one Landau level.

To determine the parity of the 4-spinor under particle interchange, we perform the following operation:\cite{Sabio:10}
\beq
\dl{z_R,z_r}\dr{\vec{\Psi}} \equiv\bem
\Psi_{AA}(z_R,z_r)\\\Psi_{AB}(z_R,z_r)\\\Psi_{BA}(z_R,z_r)\\\Psi_{BB}(z_R,z_r)\eem \rightarrow \bem \Psi_{AA}(z_R,-z_r)\\\Psi_{BA}(z_R,-z_r)\\\Psi_{AB}(z_R,-z_r)\\\Psi_{BB}(z_R,-z_r)\eem,
\eeq
where $A,B$ denote the sublattice degree of freedom. We thus see that the parity of the two-Dirac-electron eigenstates depends not only on the total relative angular momentum $(m+n)$, which fixes the exponent in the variables $z_r,\bar{z}_r$ in the wavefunction $\dl{z_r,\bar{z}_r}\dr{n,m}_r$, but also on the second and third components of the 4-spinor. From Table~I, we find that, for $\dr{\Psi_{M,m}(N=2)\,}$ with even $m$, the states (I), (III), (VI), (VII) and (VIII) are symmetric under particle exchange; whereas the states (II), (IV) and (V) are antisymmetric under particle exchange.

\subsection{Wavefunctions resulting by addition of an extra particle to the one-particle state}
In addition to calculating the eigenstates and the eigenvalues $E$ of the two-particle interacting problem, in order to compute the DOS in Eq.~(1), we also need to construct  the wavefunction resulting when a particle is added to the single-particle state $|\Psi_0(N=1)\,\rangle$, while taking into account the overall symmetry. Since the particle-addition process is assumed to be instantaneous, and not to perturb the host state, the wavefunction resulting by the addition of one particle to the one-particle state can be constructed by taking the product of two single-particle states and (anti-)symmetrizing it with the symmetrization operator $\mathcal{P}_{S}$ or antisymmetrization operator $\mathcal{P}_{AS}$. Notice that, here, we need to take into account both the symmetrization and the antisymmetrization of the orbital wavefunctions. This is because of the spin-valley component that can also be antisymmetric or symmetric, such that the total wavefunction satisfies fermion statistics.

Close to $\nu=-2$, we take the single-particle state to be the $n=0$ LL wavefunction $(0,\dr{0,m_1})$, see Eq.~(5). The addition of an extra electron in the same LL results in the two-body state
\beq
\dr{\vec{\Psi}_{S(AS)}^{\tilde{n}=0}}&=&\mathcal{P}_{S(AS)} \biggl[ \bem 0 \\ \dr{0,m_1} \eem\otimes \bem 0 \\ \dr{0,m_2} \eem\biggr]\\
&=&\f{1}{\sqrt{2}}\bem 0 \\0\\0\\\dr{0,m_1}_1\dr{0,m_2}_2\pm\dr{0,m_2}_1\dr{0,m_1}_2\eem,\nn
\eeq
where the subscripts 1 and 2 in $\dr{n,m}_1\dr{n',m'}_2$ denote the subspaces for the respective particles.

For a generalization to any integer filling $n$, it is justifiable to take the $(n-1)$ LLs to be inert. The two-body state $\dr{\vec{\Psi}_{S(AS)}^{\tilde{n}}} $ can then be constructed in a similar manner as for the $n=0$ case. For example, take the $n=1$ LL case just above the filling $\nu=2$, the instantaneous addition of an electron to the $n=1$ LL wavefunction $(\dr{0,m_1},\dr{1,m_1})/\sqrt{2}$ results in
\beq
\dr{\vec{\Psi}_{S(AS)}^{\tilde{n}=1}}&=&\f{1}{2}\mathcal{P}_{S(AS)} \biggl[ \bem \dr{0,m_1} \\ \dr{1,m_1} \eem\otimes \bem \dr{0,m_2} \\ \dr{1,m_2} \eem\biggr]\\
&=& \f{1}{2\sqrt{2}}\bem \dr{0,m_1}_1\dr{0,m_2}_2\pm\dr{0,m_2}_1\dr{0,m_1}_2 \\ \dr{0,m_1}_1\dr{1,m_2}_2 \pm\dr{0,m_2}_1\dr{1,m_1}_2 \\ \dr{1,m_1}_1\dr{0,m_2}_2\pm\dr{1,m_2}_1\dr{0,m_1}_2 \\ \dr{1,m_1}_1\dr{1,m_2}_2\pm\dr{1,m_2}_1\dr{1,m_1}_2\eem.\nn
\eeq

For a translationally invariant system, the DOS calculation does not depend on the angular momentum $m_2$ of the added particle, such that we may set $m_2=0$.\cite{MacDonald:10} In terms of ladder operators, the two wavefunctions Eq.~(13) and Eq.~(14) are then given by
\beq \label{eqn op}
\dr{\vec{\Psi}_{S(AS)}^{\tilde{n}=0}}&=&\f{1}{\sqrt{2 m_1!}} \bem 0 \\ 0\\0 \\\bigl[ (b_1^\dag)^{m_1}\pm(b_2^\dag)^{m_1}\bigr]\eem\dr{0,0}_1\dr{0,0}_2, \nn\\
\\
\dr{\vec{\Psi}_{S(AS)}^{\tilde{n}=1}}&=&\f{1}{2\sqrt{2 m_1!}} \bem \bigl[ (b_1^\dag)^{m_1}\pm(b_2^\dag)^{m_1}\bigr] \\ a_2^\dag\bigl[ (b_1^\dag)^{m_1}\pm(b_2^\dag)^{m_1} \bigr] \\a_1^\dag\bigl[ (b_1^\dag)^{m_1}\pm(b_2^\dag)^{m_1}\bigr] \\a_1^\dag a_2^\dag\bigl[ (b_1^\dag)^{m_1}\pm(b_2^\dag)^{m_1}\bigr] \eem\dr{0,0}_1\dr{0,0}_2,\nn\\
\eeq
respectively. By substituting the CM and relative coordinates operators $a_{1,2}=(a_R\pm a_r)/\sqrt{2}$, $b_{1,2}=(b_R\pm b_r)/\sqrt{2}$, and using the binomial expansion, the wavefunctions in the new basis become
\beq
\dr{\vec{\Psi}_{S(AS)}^{\tilde{n}=0}}&=&\sum_{l=0}^{m_1} F_{l,m_1}^{S(AS)}\bem 0\\ 0 \\0\\1  \eem \dr{0,l}_R\dr{0,m_1-l}_r,\\
\dr{\vec{\Psi}_{S(AS)}^{\tilde{n}=1}}&=&\f{1}{2}\sum_{l=0}^{m_1}F_{l,m_1}^{S(AS)} \bem 1\\ \f{1}{\sqrt{2}}(a_R^\dag-a_r^\dag) \\\f{1}{\sqrt{2}}(a_R^\dag+a_r^\dag)\\\f{1}{2}((a_R^\dag)^2 -(a_r^\dag)^2)  \eem \nn\\
&&\tr{\ \ \ }\times \dr{0,l}_R\dr{0,m_1-l}_r,
\eeq
with the coefficients
\beq
F_{l,m_1}^{S(AS)}\equiv [1\pm(-1)^{m_1-l}]\, \sqrt{\f{l!\,(m_1-l)!}{2\, m_1! \,2^{m_1}}} \,\binom{m_1}{l}.
\eeq
We see that the product state constructed in this way is a superposition of the two-body eigenstates of the relevant energy level listed in Table~\ref{table1}.

\subsection{Total wavefunction}
Having obtained the orbital part of the wavefunction for two Dirac particles, we now take into account the remaining internal degrees of freedom relevant for graphene. The first one is the intrinsic $1/2$ spin of the electron. The second is the valley index $K,K'$ associated to the two inequivalent Dirac cones in the graphene energy spectrum. These are separated by a large reciprocal wavevector, and even though Coulomb interaction can in principle induce inter-valley scattering, the matrix element involving atomic-scale high-momentum exchanges are typically small.\cite{Goerbig:06} Scattering between different valleys can thus be neglected in the low-energy regime, such that the valley index may be described by a pseudo-spin-$1/2$, with respect to which the Coulomb interaction is approximately SU(2)-symmetric. The total wavefunction $\dr{\Phi}$ is then a direct product of three parts:
\beq
\dr{\Phi}=\dr{\vec{\Psi}}\otimes \dr{\tr{spin, valley}}.
\eeq
The parity of the orbital part $\dr{\vec{\Psi}}$ has been discussed in Sect.~IIIA and IIIB, whereas the spin and valley parts can separately form either a singlet or a triplet state, respectively. The total two-body wavefunction must be antisymmetric under particle exchange.

\section{The DOS of graphene for $n=0$}\label{sec:4}
In this section, we compute the electron-addition DOS close to $\bar{\nu}\simeq 0$ in the $n=0$ LL, that is just above the filling $\nu=-2$ (or the electron-removal DOS close to $\nu=2$ by particle-hole symmetry). We first note that from Eq.~(13), while we fix $m_2=0$ by invoking translational invariance, the groundstate $\dr{\Psi_0(N=1)\,}$ remains macroscopically degenerate in the quantum number $m_1$. Therefore, the local DOS needs to be averaged over the $N_{B}$-fold degeneracy, e.g.,
\beq
&&A_{+}^{\tilde{n}}(\omega)= \f{1}{N_{B}}\sum_{m_1=0}^{N_{B}-1} \sum_{\alpha}\delta\bigl(\omega-E_{\alpha,0}\bigr)\nn\\
&& {\tr \ \ \ }\bigl|\langle \Psi_{\alpha}(N=2)| c_{\tilde{n},m_1,\sigma,K/K'}^{\dagger}|\Psi_{0}(N=1)\,\rangle\bigr|^2,
\eeq
for the electron-addition part of the DOS.

In the presence of a high magnetic field, the $N=1$ groundstate is a spin-polarized state with a Zeeman energy $-\Delta_z/2$, where $\Delta_z$ is the energy splitting between the majority and minority spin states. However, the groundstate electron can belong to either the $K$ or $K'$ valley. Now, when an extra electron is injected, the latter can also have a spin pointing either parallel or anti-parallel to the ground state electron, and it can reside on either the $K$ or $K'$ valley.

We first consider the groundstate electron residing on the $K$ valley and the added electron being spin parallel ($S_z=1$, where $S_z$ is the total spin component along the magnetic field direction) but belonging to either of the valleys. It then follows that the total wavefunction with an added particle can be either $\dr{\Phi}=c_{m_1,\uparrow,K}^{\dagger}|\Psi_{0,\uparrow,K}(N=1)\,\rangle$ that is
\beq
\dr{\vec{\Psi}_{AS}^{\tilde{n}=0}}\otimes\dr{\tr{spin-triplet, valley-triplet}},
\eeq
or $\dr{\Phi}=c_{m_1,\uparrow,K'}^{\dagger}|\Psi_{0,\uparrow,K}(N=1)\,\rangle$ that is given by either
\beq
&\dr{\vec{\Psi}_{S}^{\tilde{n}=0}}\otimes\dr{\tr{spin-triplet, valley-singlet}},& \nn\\
&\tr{or}&\nn\\
&\dr{\vec{\Psi}_{AS}^{\tilde{n}=0}}\otimes\dr{\tr{spin-triplet, valley-triplet}}.&
\eeq
Here, $\dr{\vec{\Psi}_{S(AS)}^{\tilde{n}=0}}$ are given by Eq.~(17). Substituting them into Eq.~(21), and summing over all $N=2$ interacting eigenstates $\dr{\Psi_{\alpha}(N=2)\,}$ (partly summarized in Table~I), we obtain
\beq
&&A_{+,S_z=1}^{\tilde{n}=0}(\omega)=\f{2}{N_{B}}\sum_{m\in even}\delta\bigl(\omega-V_m^{n=0}+\Delta_z/2\bigr)\nn\\
&&{\ \ \ \ \ \ }+\f{6}{N_{B}}\sum_{m\in odd}\delta\bigl(\omega-V_m^{n=0}+\Delta_z/2\bigr).
\eeq
This is the result which one also obtains in the case of 2D electrons in GaAs heterostructures, that is the weight of the peaks corresponding to odd pseudopotentials is 3 times larger than that for even ones. However, as we shall see, the four-component structure of graphene LLs yields eventually a different result than the two-component structure in non-relativistic LLs, when $S_z=0$ two-particle states are taken into account.

In principle, even though the eigenstate summation is performed over all two-body eigenstates, the only one yielding a non-trivial contribution is the eigenstate (I) of Table~I. We also note that there is an extra Zeeman energy cost of $-\Delta_z$ associated with the interacting eigenstate because of the spin-triplet component. Furthermore, to write down the above expression we have employed the summation formula
\beq
\sum_{m_1=0}^{N_B-1} \f{2}{2^{m_1}}\f{_1!}{(m_1-m)!m!}\,\rightarrow\, 4
\eeq
in the thermodynamic limit $N_B\rightarrow \infty$, for any integer $m$.\cite{MacDonald:10} A parallel analysis can be made for the ground state electron residing on the $K'$ valley, which leads to the same result.

On the other hand, for the addition of an electron with opposite spin, the resulting state with an identical valley $\dr{\Phi}=c_{m_1,\downarrow,K}^{\dagger}|\Psi_{0,\uparrow,K}(N=1)\,\rangle$ gives rise to either
\beq
&\dr{\vec{\Psi}_{S}^{\tilde{n}=0}}\otimes\dr{\tr{spin-singlet, valley-triplet}},&\nn\\
&\tr{or}&\nn\\
&\dr{\vec{\Psi}_{AS}^{\tilde{n}=0}}\otimes\dr{\tr{spin-triplet, valley-triplet}};&
\eeq
and the resulting state with an opposite valley $\dr{\Phi}=c_{m_1,\downarrow,K'}^{\dagger}|\Psi_{0,\uparrow,K}(N=1)\,\rangle$ gives rise to either of the states:
\beq
&\dr{\vec{\Psi}_{AS}^{\tilde{n}=0}}\otimes\dr{\tr{spin-singlet, valley-singlet}},&\nn\\
&\dr{\vec{\Psi}_{S}^{\tilde{n}=0}}\otimes\dr{\tr{spin-triplet, valley-singlet}},&\nn\\
&\dr{\vec{\Psi}_{S}^{\tilde{n}=0}}\otimes\dr{\tr{spin-singlet, valley-triplet}},&\nn\\
&\tr{\ \ or}&\nn\\
&\dr{\vec{\Psi}_{AS}^{\tilde{n}=0}}\otimes\dr{\tr{spin-triplet, valley-triplet}}.
\eeq
The resulting DOS contribution is
\beq
&&A_{+,S_z=0}^{\tilde{n}=0}(\omega)=\f{4}{N_{B}}\sum_{m\in even}\delta\bigl(\omega-V_m^{n=0}-\Delta_z/2\bigr)\nn\\
&&{\ \ \ \ \ \ }+\f{4}{N_{B}}\sum_{m\in odd}\delta\bigl(\omega-V_m^{n=0}-\Delta_z/2\bigr).
\eeq

When putting together the contributions from adding a parallel-spin electron and an opposite-spin electron to the DOS
\beq
A_{+}^{\tilde{n}=0}(\omega)&=&A_{+,S_z=0}^{\tilde{n}=0}(\omega)+A_{+,S_z=1}^{\tilde{n}=0}(\omega),
\eeq
the total weight for the odd $m$ peak becomes 5/3 times larger than that for the even $m$ (while being possible to neglect the Zeeman energy difference). 

As we have already mentioned above, the relative weight $5/3\simeq 1.67$ between the spectral weight of the odd pseudopotentials with respect to the even ones is a benchmark of the underlying four-component structure of graphene LLs, due to the spin-valley degeneracy. In the case of a two-component system, such as in a conventional 2DEG in GaAs heterostructures, the ratio would be 3.\cite{MacDonald:10} This result is retrieved in the $n=0$ graphene LL close to the charge-neutrality point ($\nu\simeq 0$), where one of the spin-valley components (say the spin component in the case of a dominant Zeeman effect) is completely frozen. The relative spectral weight between the odd and even pseudopotentials therefore yields insight into the multi-component structure of LLs.

\section{The DOS of graphene for general $n$}
\label{sec:5}
It is now straightforward to generalize our study to other LLs. For a filling factor $\nu\simeq 2$, we start by considering a groundstate which is fully occupied for all $n<1$ LLs and a single spin-polarized electron at $n=1$. The addition of an electron results in the eigenstate described in Eq.~(18). Following the same procedure to compute the DOS as in the previous section, we obtain
\beq
&&A_{+,S_z=1}^{\tilde{n}=1}(\omega)=\f{2}{N_{B}}\sum_{m\in even}\delta\bigl(\omega-\Omega_{1,m}+\Delta_z/2\bigr)\nn\\
&&{\ \ \ \ \ \ }+\f{6}{N_{B}}\sum_{m\in odd}\delta\bigl(\omega-\Omega_{1,m}+\Delta_z/2\bigr)
\eeq
and
\beq
&&A_{+,S_z=0}^{\tilde{n}=1}(\omega)=\f{4}{N_{B}}\sum_{m\in even}\delta\bigl(\omega-\Omega_{1,m}-\Delta_z/2\bigr)\nn\\
&&{\ \ \ \ \ \ }+\f{4}{N_{B}}\sum_{m\in odd}\delta\bigl(\omega-\Omega_{1,m}-\Delta_z/2\bigr).
\eeq
where now, only the eigenstate (VII) contributes to the DOS, and $\Omega_{1,m}=2\sqrt{2}\hb v/l_B +(5/8)\,V_{m}^{n=0}+ (1/4)\,V_{m}^{n=1}+ (1/8)\,V_{m}^{n=2}$. Compared to the usual 2DEG system, we see that the position of the peak does not only contain information about the $n=0$ LL pseudopotential $V_{m}^{n=0}$ but also on higher LL pseudopotentials $V_m^{n=1,2}$. This is due to the fact that a general two-body eigenstate of the interacting problem consists of spinorial components that occupy at the same time different LLs $n$ in the relative coordinate subspace.

Let us also write down the solution for the DOS of electron addition close to $\nu\simeq 6$. The two-body state for an $n=2$ LL electron with an added particle is given by
\begin{widetext}
\beq
\dr{\vec{\Psi}_{S(AS)}^{\tilde{n}=2}}&=&\f{1}{2}\sum_{l=0}^{m_1}F_{l,m_1}^{S(AS)} \bem \f{1}{2}((a_R^\dag)^2 -(a_r^\dag)^2)\\ \f{1}{4}((a_R^\dag)^3+(a _r^\dag)^3-(a_R^\dag)^2a_r^\dag -a_R^\dag (a_r^\dag)^2)\\\f{1}{4}((a_R^\dag)^3-(a _r^\dag)^3+(a_R^\dag)^2a_r^\dag -a_R^\dag (a_r^\dag)^2)\\\f{1}{8}((a_R^\dag)^4 +(a_r^\dag)^4-2 (a_R^\dag)^2 (a_r^\dag)^2 )  \eem \dr{0,l}_R\dr{0,m_1-l}_r
\eeq
\end{widetext}
and taking the overlap with the interacting eigenstate (VIII) from Table~\ref{table1}, the electron-addition parts of the DOS are given by
\beq
&&A_{+,S_z=1}^{\tilde{n}=2}(\omega)=\f{2}{ N_{B}}\sum_{m\in even}\delta\bigl(\omega-\Omega_{2,m}+\Delta_z/2\bigr)\nn\\
&&{\ \ \ \ \ \ }+\f{6}{  N_{B}}\sum_{m\in odd}\delta\bigl(\omega-\Omega_{2,m}+\Delta_z/2\bigr)
\eeq
and
\beq
&&A_{+,S_z=0}^{\tilde{n}=2}(\omega)=\f{4}{  N_{B}}\sum_{m\in even}\delta\bigl(\omega-\Omega_{2,m}-\Delta_z/2\bigr)\nn\\
&&{\ \ \ \ \ \ }+\f{4}{  N_{B}}\sum_{m\in odd}\delta\bigl(\omega-\Omega_{2,m}-\Delta_z/2\bigr).
\eeq
where $\Omega_{2,m}=4\hb v/l_B+(13/32)\,V_{m}^{n=0}+ (1/16)\,V_{m}^{n=1}+(1/4)\,V_{m}^{n=2}+(3/16)\,V_{m}^{n=3}+(3/32)\,V_{m}^{n=4}$. Thus, the DOS at this filling factor contains rich information on the pseudopotentials $V_m^{n}$ belonging to many LLs.

\begin{table}[c]
\caption{Two-body eigenvalues and eigenstates}
\begin{scriptsize}
\centering
\renewcommand{\arraystretch}{1.5}
\begin{tabular}{ |c | c | c | c|}
\hline
&$\dr{\Psi_{M,m}(N=2)}$ & $E_0/(\hb v/l_B)$ & $E-E_0$\\
\hline
\hline
(I) & $\bem 0\\0\\0\\ \dr{0,M}_R\dr{0,m}_r\eem$ & 0 & $V_{m}^{n=0}$ \\
\hline
(II) & $\bem 0\\\f{1}{2}[\dr{0,M}_R\dr{1,m}_r- \dr{1,M}_R\dr{0,m}_r] \\\f{1}{2}[\dr{0,M}_R\dr{1,m}_r+\dr{1,M}_R\dr{0,m}_r\\0\eem$  & 0 & $\f{1}{2}\bigl( V_{m}^{n=0}+V_{m}^{n=1}\bigr)$ \\
\hline
(III) & $\bem 0\\\f{1}{2}\dr{0,M}_R\dr{0,m}_r\\\f{1}{2}\dr{0,M}_R\dr{0,m}_r\\\f{1}{\sqrt{2}} \dr{1,M}_R\dr{0,m}_r\eem$ & $\sqrt{2}$ & $V_{m}^{n=0}$ \\
\hline
(IV) & $\bem 0\\\f{1}{2}\dr{0,M}_R\dr{0,m}_r\\-\f{1}{2}\dr{0,M}_R\dr{0,m}_r\\\f{1}{\sqrt{2}} \dr{0,M}_R\dr{1,m}_r\eem$ & $\sqrt{2}$ & $\f{1}{2}\bigl( V_{m}^{n=0}+V_{m}^{n=1}\bigr)$ \\
\hline
(V) & $\bem 0\\\f{1}{2\sqrt{2}}\bigl[\dr{0,M}_R\dr{1,m}_r+ \dr{1,M}_R\dr{0,m}_r\bigr] \\\f{1}{2\sqrt{2}}\bigl[\dr{0,M}_R\dr{1,m}_r-\dr{1,M}_R\dr{0,m}_r\bigr]\\\f{\sqrt{2}}{2} \dr{1,M}_R\dr{1,m}_r\eem$ & $2$ & $\f{1}{4}V_{m}^{n=0}+ \f{3}{4}V_{m}^{n=1}$ \\
\hline
(VI) & $\bem 0\\\f{1}{2\sqrt{2}}\bigl[\dr{1,M}_R\dr{0,m}_r+ \dr{0,M}_R\dr{1,m}_r\bigr] \\\f{1}{2\sqrt{2}}\bigl[\dr{1,M}_R\dr{0,m}_r-\dr{0,M}_R\dr{1,m}_r\bigr]\\ \f{1}{2}\bigl[\dr{2,M}_R\dr{0,m}_r+\dr{0,M}_R\dr{2,m}_r\bigr] \eem$ & $2$ & $\f{1}{2}V_{m}^{n=0}+\f{1}{4} V_{m}^{n=1}+\f{1}{4}V_{m}^{n=2}$\\
\hline
(VII) & $\bem \f{1}{2} \dr{0,M}_R\dr{0,m}_r\\\f{1}{2\sqrt{2}}\bigl[\dr{1,M}_R\dr{0,m}_r- \dr{0,M}_R\dr{1,m}_r\bigr]\\\f{1}{2\sqrt{2}}\bigl[\dr{1,M}_R\dr{0,m}_r+\dr{0,M}_R\dr{1,m}_r\bigr]\\ \f{1}{2\sqrt{2}}\bigl[\dr{2,M}_R\dr{0,m}_r-\dr{0,M}_R\dr{2,m}_r\bigr]\eem$ & $2 \sqrt{2}$ & $\f{5}{8}V_{m}^{n=0}+ \f{1}{4}V_{m}^{n=1}+\f{1}{8}V_{m}^{n=2}$ \\
\hline
(VIII) & $ \bem \f{\sqrt{2}}{4} \bigl[\dr{2,M}_R\dr{0,m}_r-\dr{0,M}_R\dr{2,m}_r\bigr]\\
\f{\sqrt{6}}{8}\bigl[\dr{3,M}_R\dr{0,m}_r+ \dr{0,M}_R\dr{3,m}_r\bigr]-\f{\sqrt{2}}{8}\bigl[\dr{2,M}_R\dr{1,m}_r+ \dr{1,M}_R\dr{2,m}_r\bigr]\\
\f{\sqrt{6}}{8}\bigl[\dr{3,M}_R\dr{0,m}_r- \dr{0,M}_R\dr{3,m}_r\bigr]+\f{\sqrt{2}}{8}\bigl[\dr{2,M}_R\dr{1,m}_r- \dr{1,M}_R\dr{2,m}_r\bigr]\\
\f{\sqrt{6}}{8}\bigl[\dr{4,M}_R\dr{0,m}_r+ \dr{0,M}_R\dr{4,m}_r\bigr]-\f{1}{4}\dr{2,M}_R\dr{2,m}_r\eem$ & $4$ & $\substack{\f{13}{32}V_{m}^{n=0}+ \f{1}{16}V_{m}^{n=1}+\f{1}{4}V_{m}^{n=2}\\ \\
+\f{3}{16}V_{m}^{n=3}+\f{3}{32}V_{m}^{n=4}}$
\\
\hline
\end{tabular}
\label{table1}
\end{scriptsize}
\end{table}

\section{Discussions and conclusions}
\label{sec:6}

We have calculated the tunneling DOS in graphene in high magnetic fields when the filling factor is close
to $\nu_n\pm 2(2n+1)$. In order to describe the electronic interactions, we have used the method of Haldane's pseudopotentials, which describes the two-particle interacting eigenstates of a system in strong magnetic field. The method is valid for a system in the very close proximity of a completely filled LL, such that, besides an integer number of filled and inert LLs, only a single electron or hole are present. Although this limit may seem, at first sight, extremely theoretical, it describes
the experimental situation of a very sparsely electron- or hole-filled LL, in which the average distance between the particles is larger than the cyclotron radius $R_C=l_B\sqrt{2n+\delta_{n,0}}$ in graphene, that is $\bar{\nu}\ll 1/(2n+\delta_{n,0})$. The tunneling from the STM tip, which injects a second particle into the system, can thus measure the overlap between the resulting state and the two-particle interacting eigenfunctions of the system. Also, it allows one to measure the difference in energy between the two-particle interacting states of the system and the one-particle state, thus yielding information about the strength of the interactions.

Our calculations revealed that the DOS spectrum exhibit peaks, the  energy of which can be related directly to the energies of Haldane's pseudopotentials. While the $n=0$ state is quite similar to the $n=0$ LL in non-relativistic 2D electron systems with a parabolic band dispersion, the higher LL DOS structures are different, in that the energies in the spectrum do not result from a single pseudopotential value, but involve combinations of these values, corresponding to  states with different angular momenta. This is a direct consequence of two graphene-specific properties. The first one is that the spinorial eigenstates have (sublattice) components
in different non-relativistic LLs. The second one is that, as a consequence of the Lorentz invariance of the underlying Dirac equation, the
center of mass and the relative degrees of freedom are intimately coupled. Finally, the relative spectral weight between the peaks corresponding to odd and even pseudopotentials yields insight into the multi-component structure of graphene LLs.

It would be interesting to generalize our results for larger partial fillings, eventually moving towards the regime of the fractional quantum Hall effect. Such an analysis would allow one to make predictions about the experimental spectroscopic signature of different highly-delicate quantum Hall states, such as the $\nu=1/2$ state, that do not have distinct anomalous transport signatures; understanding the nature of such states has been a long-standing question in the study of the FQHE. While the fractional quantum Hall effect has been measured only recently in graphene in transport experiments,
spectroscopic measurements may yield additional information about the relevant electronic interactions in graphene LLs and thus about the nature of strongly-correlated phases in partially filled levels.

\acknowledgements

This work was supported by the ANR project NANOSIM GRAPHENE under Grant No. ANR-09-NANO-016, and by the FP7 ERC Starting Independent Researcher Grant NANO-GRAPHENE 256965.

\end{document}